\documentclass[]{aa}
\usepackage{graphicx}
\begin{document}


\title{
Post-outburst radio observation of the region around McNeil's nebula
(V1647 Ori)}

\author{S. Vig$^1$, S.K. Ghosh$^1$, V.K. Kulkarni$^2$ \and D.K. Ojha$^1$}

 \offprints{S. Vig, \email{sarita@tifr.res.in}}

\institute{$^1$Tata Institute of Fundamental Research, Mumbai 400 005, India\\
$^2$National Centre for Radio Astrophysics, Pune 411 007, India \\ }

\date{Accepted 10 Oct 2005}

\abstract{
We present post-outburst ($\sim$ 100 days after outburst) radio continuum 
observation of the region ($\sim 30'\times 
30'$) around McNeil's nebula (V1647 Orionis). The observations were carried 
out using the Giant Metrewave Radio Telescope (GMRT), India, at 1272 MHz on 
2004 Feb 14.5 UT. Although 8 sources have been detected within a 
circular diameter of $25'$ centred on V1647 Ori, we did not detect any radio 
continuum emission from McNeil's nebula. We assign a $5\sigma$ upper limit of 
0.15 mJy/beam for V1647 Ori where the beam size is $5.6''\times2.7''$. 
Even at higher frequencies of 4.9 and 8.5 GHz (VLA archival data), no radio 
emission has been detected from this region.
 Three scenarios namely, emission from homogeneous HII region, ionised 
stellar wind and shock ionised gas, are explored in the light of our GMRT 
upper-limit. For the case of homogeneous HII region, the radius of the 
emitting region is 
constrained to be $\la 26$ AU corresponding to a temperature $\ga 2,500$ K, 
which is consistent with the reported radio and H$\alpha$ emission. 
In the ionised stellar wind picture, our upper limit of radio emission 
translates to   
$\dot{M}/v_{\infty}< 1.2-1.8\times10^{-10} M_{\odot} yr^{-1} km^{-1} s$.
 On the other hand, if the stellar wind shocks the  
dense neutral (molecular) cloud, the radio upper limit implies that the 
fraction of the wind encountering the dense obstacle is $<50$\%. Based on a 
recent measurement of X-ray outburst and later monitoring, the expected radio 
emission has been estimated. Using our radio limit, the radius ($\la36$ AU) and 
electron density ($\ga 7.2\times10^7$ cm$^{-3}$) of the radio 
emitting plasma 
have been constrained using a two phase medium in pressure equilibrium 
for a volume filling factor of 0.9. 

}

\titlerunning{Radio observations of McNeil's nebula}
\authorrunning{S. Vig, et al.}
\maketitle


\section{Introduction}
J. W. McNeil, in January 2004, reported the presence of a new reflection nebula,
now known as McNeil's nebula, in the Lynds dark cloud 1630 in Orion B molecular
complex (McNeil, 2004). 
McNeil's nebula (V1647 Orionis) is at a distance of 400 pc (Anthony-Twarog, 1982) and is positionally 
coincident to the infrared source IRAS 05436-0007. Reipurth \& Aspin (2004) 
presented the optical and near IR photometry of the source V1647 Ori (on 2004 
Feb 14 and Feb 3 UT, respectively), and 
found that the star brightened by about 3 mag in the near infrared (J, H, K) 
with respect to 2MASS measurements (1998). The P-Cygni profile of the
H$\alpha$ line emission from the region around this source implies 
considerable mass loss in powerful winds with velocities reaching upto $\sim 
600$ km s$^{-1}$(Reipurth \& Aspin, 2004). Brice\~no et al. (2004) constrained 
the epoch of the onset of the outburst to be within 
28 Oct - 15 Nov, 2003 by means of I band images of this region. Optical and near
 infrared photometry of the exciting star, V1647 Ori, show that it is 
 variable atop a general decline in brightness of about 0.3-0.4 mag in 87 days,
 between 2004, February 10 and May 7 (Walter et al., 2004). Vacca et al. (2004) 
have presented the near infrared spectroscopy at medium resolution of the 
McNeil's nebular object after the outburst on March 9, 2004 UT. They suggest 
it to be a low mass Class I protostar with mass outflow ($4\times10^{-8} 
M_{\odot}/yr$) occuring in a dense ionised wind of velocity $\sim 400$ km 
s$^{-1}$. Andrews et al. (2004) show that the 12 $\mu$m flux of this source 
has increased by a factor of $\sim25$ after the outburst whereas the 
submillimeter continuum is at the pre-outburst level. 

 Kastner et al. (2004) observed a surge in X-ray brightness coincident with 
the optical and near infrared eruption, and suggest that it could be due to a 
sudden onset of rapid accretion phase. More recent X-ray observations by Grosso
 et al. (2005) have led to detection of X-ray variability within a $\sim12$ 
hour duration (2004 April 4). They conclude the soft X-rays to originate from 
accretion shock onto the photosphere of a low mass star and the harder 
component due to magnetic reconnection events. Enhanced accretion events from a 
circumstellar accretion disk are believed to be responsible for such outbursts 
in young stars. EX Lupi - like objects (EXors) and FU Orionis objects (FUors) 
are classes of optically variable young stellar objects associated with 
eruptions. EXor eruptions are generally shorter (few months to a few years) as 
compared to FUor brightenings (which last many decades).
While Walter et al. (2004) suggest that the McNeil's nebula is an FUor like
outburst, McGehee et al. (2004) from optical and near infrared measurements 
infer that the brightening is suggestive of EXor like outburst. \'Abrah\'am et 
al. (2004) compiled the infrared-submm-mm spectral energy distribution of
the outburst star in its quiescent phase and find that it resembles those of
FUor objects indicating the presence of circumstellar envelope. Ojha et 
al. (2005) have presented the short time variability as well as near 
infrared morphology of the McNeil's nebula and showed a significant variation 
in the source brightness ($>0.15$ mag) within a period of one week. More 
recently, Rettig et al. (2005) have presented high resolution infrared spectra 
of V1647 Ori. They suggest that the CO emission lines are likely to originate 
from $\sim2,500$ K gas in the inner accretion disk region.

Young stellar objects in their pre-main sequence stage
generally have very weak radio emission. Although Br$\gamma$ and H$\alpha$
lines, which trace the hot ionised gas, have already been detected in McNeil's 
nebula, no radio study exists in literature. Hence, we carried out radio 
observations of McNeil's nebula on 2004
 Feb 14 UT ($\sim22$ days after the first announcement by McNeil on 2004 Jan 
23 UT; $\sim 91-109$ days after the outburst as inferred by Brice\~no et al., 
2004) using the Target of Opportunity (ToO) allocation of the observing time 
on the Giant Metrewave Radio Telescope (GMRT). In this paper, we present the 
radio measurement of the region around McNeil's nebula, carried out at 1272 
MHz. We explore three possible scenarios to understand the radio emission 
each involving thermal bremstrahlung from the ionised gas. 
Section 2 describes the observations and data reduction. While Section 3 gives 
the results; in Section 4, radio emission from three scenarios, \textit{viz}. 
homogeneous HII region, ionised stellar wind and shock ionised region 
have been considered. Physical constraints on 
the emitting region based on radio and other available measurements (H$\alpha$ 
and X-ray) have also been discussed in this section. 
In Section 5, we present our conclusions.

\section{Observations and Data reduction}

The radio continuum interferometric observations of V1647 Ori were carried out 
at 1272 MHz on 2004, Feb 14.5 UT using the GMRT located at Khodad 
(19$^{\circ}$ $5'$ $30''$ N, 74$^{\circ}$ $3'$ $0''$ E) in India. GMRT has a 
``Y'' shaped hybrid configuration of 30 antennas, each antenna of 45-m 
diameter. Of the 30 antennas, six each are placed along the east, west 
and south arms 
(arm length $\sim 14$ km). The longest interferometric baseline among the most 
distant antennas is about 25 km. The remaining twelve are located randomly 
in a compact $1\times1$ km$^2$ at the center (Swarup et al., 1991).
The phase center of the observations was ($\alpha$,$\delta$)$_{2000}$ = ($05^h$ 
$46^m$ $14.1^s$, -00$^{\circ}$ 05$'$ 48$''$), close to the centre of the 
diffuse 
optical emission (Reipurth \& Aspin, 2004). The total  duration of the 
observations was 3 hours. The standard flux calibrator 3C48 was observed at 
the beginning and end of the observations.
 The QSO B0500+019 was used for phase calibration. The continuum
observations were carried out with 16 MHz bandwidth.

NRAO Astronomical Image Processing System (AIPS) was used for the reduction of 
data. Before analysing the data, it was carefully
checked for radio interference and other instrumental problems and the 
corrupted data were
removed. The amplitude and phase calibrated data were Fourier transformed to
make images which are then deconvolved using the IMAGR task in AIPS. This task
 uses the CLEAN algorithm. The sky was approximated by a 2-dimensional
plane ($\sim 8'$ from phase center) close to the phase tracking center.
However, since a region of $\sim 30'\times 30'$ has been considered, a number
of such 2-dimensional fields have been considered for the Fourier
transformation and deconvolution. These fields were projected to a single 
tangent plane on the phase center using the AIPS task FLATN. Self calibration 
was carried out using 
the task CALIB to fix the final phases and obtain the improved final maps. 
Considering that a region larger than the primary beam ($\sim 21'$) has been 
cleaned, the task PBCOR has 
been used to apply the primary beam correction to the image. For the sources 
detected in the radio map, the peak position as well as the flux densities 
(peak and integrated) were obtained using the task JMFITS in AIPS.

We have also analysed the data for V1647 Ori at 4.9 and 8.5 GHz taken 
from the NRAO Data Archive\footnote{The National Radio Astronomy Observatory is
 a facility of the National Science Foundation operated under cooperative 
agreement by Associated Universities, Inc.} obtained using the Very Large 
Array (VLA) in the B/C configuration. The observation was carried out on March
05, 2005. Since the VLA datasets did not include any of the standard flux 
calibrators, the VLA phase calibrator, 0541-056, itself was used for both the 
phase as well as flux calibration. A method of analysis, in AIPS, similar to 
that described above was used for the reduction of this data.

\section{Results}
A map of nearly $30'\times30'$ region at 1272 MHz has been generated, centred 
on V1647 Ori. 
Emission from a part of this region ($12'\times12'$) around V1647 Ori is shown 
in Fig 1. The position of V1647 Ori is marked by a `$+$' symbol. 
The synthesized beam size is $5.6''\times2.7''$ with position angle = -- 
48.9$^{\circ}$ (FWHM along the major axis corresponds to $2234$ AU for $d=400$ 
pc). The rms noise in our map is $\sim30$ $\mu$Jy/beam.
No radio emission has been detected in our map at the location of V1647 Ori or 
around ($\sim 2'$) it. We place a fairly strict 5$\sigma$ upper 
limit of $0.15$ mJy/beam for the radio emission from this source. Away from 
the McNeil's nebular region, eight radio sources were clearly detected in the 
map. The coordinates as well as the integrated flux densities of these sources 
are presented in Table 1.  
The integrated flux densities of the sources are in the range 0.94 - 8.42 
mJy. At this bright level, there are several sources detected by the NRAO VLA 
Sky Survey (NVSS) at 1400 MHz, which have also been listed in Table 1. Seven 
of the eight sources detected by us in a circular region of diameter 25$'$ 
appear in the NVSS catalog. The faintest one among the sources detected in our 
image, labelled by identification number 4 in Fig 1, does not appear in the 
NVSS catalog. 
The separations in the positions of these 
sources between our GMRT map and the NVSS catalog are in the range 
of 1.8$''$ - 5.2$''$ 
which is reasonable considering the larger 
synthesized beam for NVSS ($\sim 45''$). The identification (ID) 
numbers of the sources (Table 1) have been used to label them in Fig 1. 

Maps of the region around V1647 Ori at 4.9 and 8.5 GHz were generated 
using the VLA data. The rms noise in the maps were 25 and 18 $\mu$Jy/beam 
and the synthesized beam sizes are $4.8''\times4.0''$ and $6.6''\times2.3''$ 
at 4.9 and 8.5 GHz, respectively. No radio emission has been detected at 
the position of V1647 Ori in these maps at either frequency and we 
place 5$\sigma$ upper limits of 125 and 90 $\mu$Jy/beam at 4.9 and 8.5 GHz, 
respectively. In what follows, the limit obtained from GMRT measurements has  
been used to quantitatively constrain the physical parameters of the region 
around V1647 Ori. The upper limits at 4.9 and 8.5 GHz could be uncertain by a 
factor upto $\sim1.3$ and $2$, respectively, due to variability of the 
calibrator (Tornikoski et al. 2001). In spite of this, the limits at higher 
frequencies from VLA would also support similar conclusions based on limits at 
1272 MHz.

\section{Discussion}
Next, we explore possible scenarios and discuss the implications of
our upper limit on radio emission from V1647 Ori. It is believed that the 
source of outburst, V1647 Ori, is a very young low mass stellar object
(Vacca et al., 2004; \'Abrah\'am et al., 2004). 
In the literature, a number of 
suggestions have been made about radio emission from embedded 
low-mass young stellar objects (Gibb, 1999).
 
The radio emission could arise from (a) shocks in 
Herbig-Haro flows (Curiel et al., 1993), (b) ionised stellar wind/jet (Martin, 
1996) (c) ionised region surrounding a protostellar accretion shock 
(Neufield \& Hollenbach, 1996), or (d) magnetic fields close to the star 
giving rise to star-disk interaction (Andr\'e, 1996).
For V1647 Ori, we explore the possibilities of thermal radio emission from 
homogeneous HII region, ionised stellar winds and shocks by outflows.

\subsection{Homogeneous HII region}
Here, we consider radio emission from a homogeneous HII region 
around the young stellar object V1647 Ori. In this scenario, we assume that 
the radio emitting region is (i) spherically symmetric and of constant density,
 (ii) pure hydrogen and dust free, (iii) isothermal, (iv) identical to the 
H$\alpha$ emitting region and optically thin in H$\alpha$ (case A of 
Osterbrock, 1989). The measured H$\alpha$ emission and the radio upper limit 
are used to constrain physical parameters like density and size of the 
emitting region for an assumed temperature. In the case of a low-mass stellar 
object like V1647 Ori, a possible source of ionising radiation (ultraviolet 
photons) could be from the boundary layer between the stellar surface and the 
accretion disk (Hartmann, 1998).

Using the normalised H$\alpha$ profile of V1647 Ori obtained by Walters et al. 
(2004) on Feb 13, 2004, from a slit whose smaller side corresponds to 600 AU 
($d=400$ pc) and their calibrated
photometry in the B, V, R and I bands, we obtain the peak of the H$\alpha$
emission to be $9.7\times10^{-15}$ erg cm$^{-2}$ s$^{-1} \AA^{-1}$. The total
integrated and dereddened H$\alpha$ emission from the slit using 
$A_V=11$ (from Vacca et al., 2004) and the extinction law of Rieke \& Lebofsky
(1985) is 4.2$\times 10^{-10}$ erg cm$^{-2}$ s$^{-1}$. According to Retigg et 
al. (2005), the temperature of the gas in the inner accretion disk of V1647 
Ori is $\sim2,500$ K based on the CO emission lines in the infrared. Vacca et 
al. (2004) interpret 
the Brackett transitions to be emitted from a gas of temperature 10,000 K. We, 
therefore, consider temperatures of 2,500 K, 5,000 K and 10,000 K for our 
discussion here. For various radii of the emitting region, we determine the 
electron densities consistent with the measured H$\alpha$ emission  
using the recombination coefficients given by Osterbrock (1989). 
These electron densities have been used to determine the total radio 
flux at 1272 MHz using the formulation of Spitzer (1978). 

Fig 2 shows the relation between electron density and various radii of the 
emitting region (in AU). These curves, labelled as `H$\alpha$ 
and Radio' in the figure, are presented for three temperatures  $2,500$ K - 
$10,000$ K. The solid line intersecting the curves for different temperatures 
represents the radio flux limit of 150 $\mu$Jy. The region below this line and 
along the curves represents the allowed parameter space. Here, for a given 
temperature, the pair of radius and electron density gives rise to consistent 
values of H$\alpha$ emission and radio upper limit. The solid line corresponds 
to an emitting region (radius) of 13-26 AU for 10,000 K - 2,500 K. For the 
same temperature range, the electron densities vary as $3.5\times10^7$ 
cm$^{-3} - 6.5\times10^6$ cm$^{-3}$, respectively.

\subsection{Ionised stellar wind}

We next explore radio emission from an ionised stellar wind. There is an 
indication of strong stellar winds from the outburst source V1647 Ori  
(P-Cygni profiles observed in the H$\alpha$ line; Vacca et al. 2004). The 
radio emission 
from ionised stellar winds of young stellar objects has been discussed by 
Panagia (1991). Under the simplistic assumptions that the stellar wind is (a) 
fully ionised, (b) spherically symmetric with constant mass-loss rate and 
velocity, (c) isothermal with the temperature of stellar wind being the same as 
that of stellar photosphere, 
and (d) is in quasi-LTE conditions, the monochromatic radio flux density is 
given by (Panagia \& Felli, 1975)

$$\frac{S_{\nu}}{\mathrm{mJy}} = 5.12 \: \left[\frac{\nu}{\mathrm{10 \:GHz}}\right]^{0.6}\: \left[\frac{T}{10^4\: K}\right]^{0.1}\: \left[\frac{\dot{M}/10^{-6}\:\frac{M_{\odot}}{yr}}{v_{\infty}/\mathrm{100 \:km\: s^{-1}}}\right]^{4/3}$$
\hspace*{-1in}
$$\times \left[\frac{d}{\mathrm{kpc}}\right]^{-2}$$  

\noindent
where $S_{\nu}$ is the expected radio flux density at frequency $\nu$ from the 
ionised stellar wind, $T$ is the temperature, $\dot{M}$ is the mass-loss rate, 
$v_{\infty}$ is the terminal velocity of the stellar wind, and $d$ is the 
distance to the source. Temperatures of isothermal stellar winds vary between 
$3\times10^3$ K to $3\times10^5$ K for stars of mass 1$M_{\odot}$ (Lamers and 
Cassinelli, 1999). For this temperature range, our upper limit of radio 
emission implies the ratio of mass-loss rate to velocity of the stellar wind, 
($\dot{M}/v_{\infty}$), to be $< 1.2-1.8\times10^{-10} M_{\odot} yr^{-1} 
km^{-1} s$. Using the inferred mass-loss rate ($\sim 4 \times 10^{-8} M_{\odot}
 yr^{-1}$) and wind velocity 
($\sim 400$ km s$^{-1}$) from the P-Cygni profile of the H${\alpha}$ emission 
by Vacca et al., (2004), we find that $\dot{M}/v_{\infty} = 1\times10^{-10} 
M_{\odot} yr^{-1} km^{-1} s$, which is consistent with our upper limit.

In order to explain the H$\alpha$ measurement (Section 4.1) for $T=10,000$ K 
and using the density determined from the mass-loss rate of a spherical
stellar wind close to the stellar surface for a 1$R_{\odot}$ star, the 
emission should be in the form of an extremely narrow bipolar flow of solid 
angle $4\times10^{-10}$ sr, which appears unphysical.

\subsection{Shock-ionised Gas}

We consider shocks by outflows as a possible source of radio emission from 
V1647 Ori although the near infrared (H$_2$) and optical ([SII]) studies by 
Reipurth \& Aspin (2004) and Vacca et al. (2004) showed no evidence of shocked 
gas. In addition, the star associated with McNeil's nebula was proposed as the 
exciting source of HH23 (Eisl\"offel \& Mundt 1997; Kastner et al. 2004). 
Shocks could form due to a supersonic stellar wind impinging upon a dense clump
 of gas in the flow close to the exciting source 
(Hartigan et al. 1987). Fast shocks (shock velocity $>60$ km s$^{-1}$) due to  
winds from young stellar objects into molecular clouds are a potential source 
of free-free radio emission since the shock energy ionises the compressed gas.
If a stellar wind with a mass-loss rate of $\dot{M}$ and terminal velocity 
$v_{\infty}$ is shocked due to the presence of
 an obstacle such as a molecular cloud, then the radio continuum flux density, 
$S_{\nu}$, in the optically thin limit is given by (Curiel et al., 1989) 

$$\frac{S_{\nu}}{\mathrm{mJy}}=3.98\times10^{-2}\;\eta\left[\frac{\dot{M}}{10^{-7}M_{\odot}yr^{-1}}\right]\left[\frac{v_{\infty}}{100 \:\mathrm{km\:s^{-1}}}\right]^{0.68} $$
$$\times \left[
\frac{d}{\mathrm{kpc}}\right]^{-2} \left[\frac{T}{10^4\:\mathrm{K}}\right]^{0.45} \left[\frac{\nu}{5\:\mathrm{GHz}}\right]^{-0.1}$$

\noindent
where $\eta$ represents the geometrical fraction 
of the stellar wind that is shocked,
 $d$ is the distance to the source, $T$ is the temperature and $\nu$ is the 
frequency. For a temperature of $10,000$ K, and using the values of distance, 
mass-loss rate and velocity of stellar wind of V1647 Ori given in Section 4.2, 
we find that the expected radio flux is $300\eta\; \mu$Jy. Hence, our limit 
implies that if the shocked ionised gas is responsible for the radio emission, 
the fraction of the wind encountering the dense molecular 
obstacle is $\eta<50$\%.

\subsection{X-ray measurements and its implications}
 Recently, an X-ray outburst has been detected from
 V1647 Ori, which illuminates the McNeil's nebula,
 by Kastner et al. (2004) based on their
 observations using the {\it Chandra} satellite. Their reported
 observations correspond to three epochs (14 Nov. 2002,
 7 March 2004 and 22 March 2004) showing quiescent (LOW),
 outburst (HIGH) and post-outburst emission.
 Thereafter Grosso et al., 2005, have detected enhanced
 X-ray variability from a long integration using the
 {\it XMM-Newton} satellite on 4 April 2004.
 Kastner et al (2004) have modelled their observations
 and concluded that the plasma responsible for the X-ray
 emission during the HIGH state is characterised by a temperature
$T_X \sim 5.6 \times 10^7$ K
 and X-ray luminosity $L_X \sim 10^{31}$ erg s$^{-1}$.
 The corresponding values for $T_X$ and $L_X$ during the LOW state are 
 $\sim 10^7$ K and $ \sim 3.0 \times 10^{29}$ erg s$^{-1}$ respectively.
 The {\it XMM-Newton} data imply slightly different values
 ($T_{hard} \sim 4.94 \times 10^7$ K,  $T_{soft} \sim 
1.06 \times 10^7$ K, $L_X \sim 1-3 \times10^{31}$ erg s$^{-1}$; 
Grosso et al.). They propose that while the soft X-rays could originate from 
accretion shock onto the photosphere of a 
low mass star, the harder component is likely to be due to magnetic 
reconnection events. Here, we consider the implications on expected radio 
emission at 1272 MHz based on the {\it Chandra} measurements.

  The X-ray emission from hot plasmas are well quantified
 in terms of ``band cooling coefficients" in different energy
 ranges as a function of temperature. We have used the
  band cooling coefficients calculated by Raymond, Cox \& Smith
 (1976). Considering the {\it Chandra} X-ray band to be
 0.5--8 keV, we obtain the emission integral, $n_e^2V$ 
(where $n_e$ is the 
electron density and $V$ is the volume occupied by the plasma),
to be $8.6 \times 10^{52}$ cm$^{-3}$ during the LOW state and
 $8.3 \times 10^{53}$ cm$^{-3}$during the HIGH state.
 Assuming a spherical emitting region of constant density,
 the above values constrain the radius, $R$ and $n_e$ for the HIGH state as 
shown by a dash-dotted line in Fig 2.

  We estimate the expected radio emission from the X-ray
 emitting plasma, assuming it to be optically thin, following
 Spitzer (1978). They are 0.01 and 0.05 $\mu$Jy corresponding to the 
 LOW and  HIGH states respectively (for $d = 400$ pc).
 These values are much smaller than (and hence consistent with) our upper limit
 at 1272 MHz. 
In case the assumption of optical thinness does not hold, then
 the emission would be even lower.

   Next, we consider the hot plasma to be clumpy
 with two components in pressure equilibrium, the hotter
 component (electron density $n_1$, temperature $T_1$)
 responsible for the X-ray emission and the other
 ($n_2$, $T_2$) component for the radio. 
The pressure equilibrium implies $ n_1T_1 = n_2T_2 $. 
 Let the volume filling factor for the hotter component be $ f = (V_{X-ray} / (V_{X-ray} + V_{radio})) $.  If $f << 1$, then
$ f \sim (V_{X-ray} /  V_{radio}) $. 
 Assuming $T_2$ = 5,000 K, we estimate the $n_e^2 V$ for the radio
 emitting gas corresponding to both the states, which we translate
 to radio flux density for optically thin case. The values turn
 out to be rather large, viz., $(2.0/f)$ and $(607/f)$ Jy corresponding to LOW
 and HIGH states respectively. Both are much larger than our limit of 150 
 $\mu$Jy. If the non-detection of V1647 Ori at 1272 MHz is purely due to
 optical depth effects, we can constrain the size, $R$, and density, $n_e$, 
 of the radio emitting phase using the upper limit in this two phase plasma 
 scenario. We find $R \la 36$ AU and $n_e \ga 7.2 \times 10^7$ cm$^{-3}$ 
 ($1.3 \times 10^9$ cm$^{-3}$) for LOW (HIGH) state corresponding to a volume 
 filling factor, $f=0.1$, 
which is depicted in Fig 2 by the striped region.

In ionised stellar wind picture, the X-ray emission considering that (i) the 
X-ray luminosity is dominated by the free-free emission of hydrogen (ii) 
$T=10,000$ K, and (iii) density determined from mass-loss rate of a spherical 
stellar wind close to the stellar surface for a 1$R_{\odot}$ star, is 
$< 10^{24}$ erg s$^{-1}$; much lower than that measured by $Chandra$. 

In Section 4.3, we have considered shocked gas as a source of
radio emission. Here, we extract physical parameters of this scenario from the 
X-ray measurements using the prescription of Raga et al. (2002).
In this scheme, a bow shock is created at a thin surface near the apex of
a dense, approximately spherical obstacle of radius $r_b$ which emits X-ray
of luminosity, $L_X$, given by

$$\frac{L_X}{L_{\odot}} = 4.1\times10^{-6}\left(\frac{n_o}{100\;\mathrm{cm^{-3}}}\right)\left( \frac{r_b}{10^{16}\;\mathrm{cm}}\right)^{2}\left(\frac{v_{s}}{100\; \mathrm{km\;s^{-1}}}\right)^{5.5}$$

\noindent
Here, $n_o$ is the pre-shock number density and $v_s$ is the shock velocity.
Taking the X-ray luminosity of V1647 Ori from $Chandra$ measurements (Section
4.4) and $v_s = 400$ km s$^{-1}$, we obtain the size of the X-ray emitting
shocked region as a function of $n_o$. This variation of density with 
$r_b$ for the LOW and HIGH states is shown in Fig 3.

\section{Conclusions}

In the present study, three scenarios have been explored to explain the radio, 
H$\alpha$ and X-ray measurements of McNeil's nebula. For the case of 
homogeneous HII region, the size of the emitting region and the electron 
density must lie in the dark-shaded region of Fig 2 ($2,500 - 10,000$ K) in 
order to be consistent with all the three measurements. A typical 
solution would be for T$=5,000$ K and $n_e=7.2\times10^7$ cm$^{-3}$; the size 
and radio flux are then given by $6.3$ AU and 19 $\mu$Jy, respectively. 
In the ionised stellar wind picture, the expected X-ray emission 
is very small. The H$\alpha$ 
measurement also cannot be explained in this scenario. The shocked gas 
picture can be consistent with X-ray measurements provided the pre-shock 
number density and the radius of the dense spherical obstacle fall near the 
curves shown in Fig 3. However, for these parameters, the H$\alpha$ 
measurement cannot be explained.

Although we have considered extremely simple pictures under 
idealized assumptions, we conclude that the homogeneous HII region scenario 
explains the radio, H$\alpha$ as well as X-ray measurements consistently while 
the ionised stellar wind and shocked gas pictures do not. For the latter two 
scenarios, we cannot rule out the possibility that these emissions 
(radio, X-ray and H$\alpha$) are at different length scales. However, in that 
case, the physical connection implied by the outburst and time variability in 
all three wavebands will remain unexplained. 

\begin{acknowledgements}
We thank the anonymous referee for many useful suggestions which have improved 
the paper. We thank the Centre Director, NCRA-TIFR and members of the 
Target of Opportunity (ToO) 
time allotment committee for making time available for our observations. We 
thank the staff of the GMRT who have made the radio observations possible. 
GMRT is run by the National Centre for Radio Astrophysics of the Tata 
Institute of Fundamental Research.
\end{acknowledgements}


\begin{table*}
\begin{center}
\caption{Radio sources detected in a circular region of diameter $25'$ around
McNeil's nebula}
\vspace{0.5cm}
\begin{tabular}{|c | c c c | c c c | c c|}
\hline
\multicolumn{1}{|c|}{} & \multicolumn{3}{c|}{GMRT} & \multicolumn{3}{c|}{NVSS} & \multicolumn{2}{c|}{Separation$^a$} \\ \hline
ID no. & RA (J2000) & Dec(2000) & Int flux &  RA(J2000) & Dec (J2000) & Int flux & $\Delta\alpha$ & $\Delta\delta$ \\
      & h  m  s & $^{\circ}$  $'$  $''$ & mJy & h  m  s & $^{\circ}$  $'$  $''$ & mJy &  $''$ & $''$ \\
\hline
1 & 05 45 27.26 & -00 05 47.7 & 2.55 & 05 45 27.46 & -00 05 47.6 & 7.5  & -3.0 & -0.1\\
2 & 05 45 39.42 &  00 05 00.2 & 8.41 & 05 45 39.64 & +00 04 58.1 & 23.9 & -3.3 & 2.1 \\
3 & 05 45 43.93 & -00 13 56.5 & 2.24 & 05 45 43.94 & -00 14 00.1 & 2.9  & -0.15 & 3.6\\
4 & 05 45 51.72 & -00 08 03.0 & 0.94 & - & - & - & - & - \\
5 & 05 45 57.71 & -00 09 27.8 & 1.24 & 05 45 57.49 & -00 09 24.5 & 2.6  & 3.3 & -3.3 \\
6 & 05 46 29.53 & -00 19 41.4 & 4.88 & 05 46 29.80 & -00 19 44.4 & 15.5 & -4.1 & 3.0 \\
7 & 05 46 40.76 & -00 18 23.8 & 8.42 & 05 46 40.80 & -00 18 25.5 & 12.0 & -0.6 & 1.7 \\
8 & 05 46 43.98 &  00 01 41.7 & 4.59 & 05 46 44.07 & +00 01 44.8 & 5.9  & -1.4 & -3.1 \\
\hline
\end{tabular}
\end{center}
$^a$GMRT - NVSS
\end{table*}



\begin {figure*}
\begin {center}
\includegraphics[height=15.0cm]{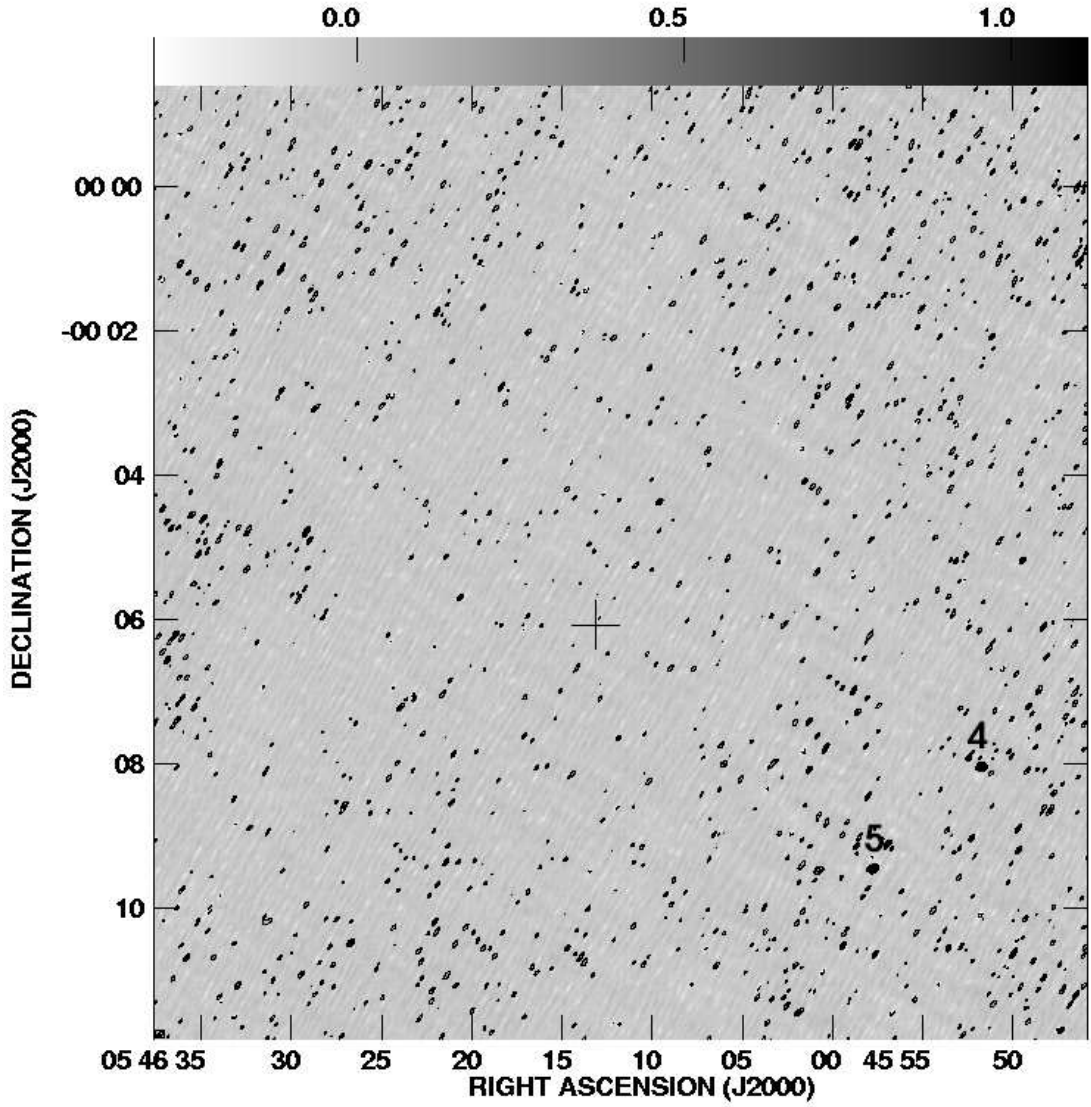}
\caption[]{Radio continuum emission at 1272 MHz from the region around V1647 
Ori (McNeil's nebula) and sources with ID 4 and 5. The position marked by a 
`$+$' symbol represents the location of V1647 Ori [$\alpha_{2000}$, $\delta_{2000} = (05^h 46^m 13.14^s, -0^{\circ} 06' 04.8'')$]. The synthetic beam size 
is $5.6''\times2.7''$ and the rms noise in the map is $\sim 30$ $\mu$Jy/beam. 
The contour levels are at $90\times(-2, 1, 2, 3, 4, 5, 6)$ $\mu$Jy/beam.}
\end{center}
\end {figure*}

\begin {figure*}
\begin {center}
\hskip -3mm
\includegraphics[height=7.0cm]{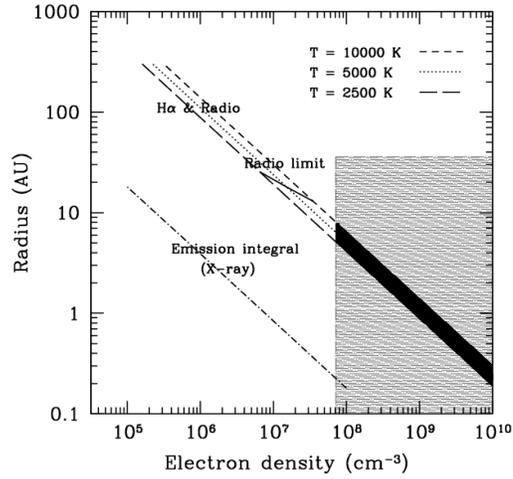}
\caption[]{The radius of emitting region as a function of electron density are 
plotted for emission from a homogeneous HII region consistent with radio and 
H$\alpha$ measurements (for 2,500, 5,000 and 10,000 K). The solid line 
corresponds to the radio upper limit of 0.15 mJy. The dash-dotted line 
represents the constraints from X-ray emission integral, $n_e^2V$, 
(HIGH state; see Section 4.4). The striped region represents the allowed 
values when the X-ray and radio emitting regions are in a two-phase plasma 
model in pressure equilibrium for a volume filling factor of 0.1 for the hotter 
plasma. The dark region includes the H$\alpha$ constraints also.}
\end{center}
\end {figure*}

\begin {figure*}
\begin {center}
\includegraphics[height=7.0cm]{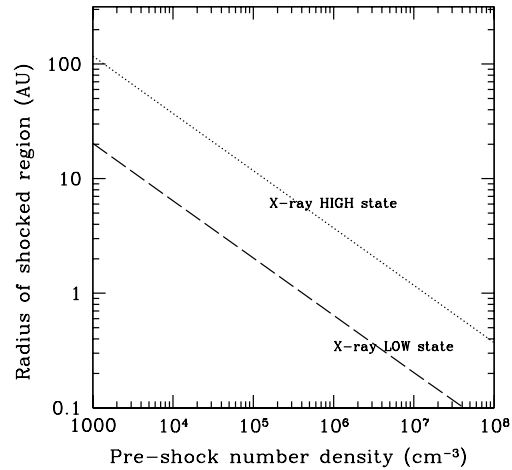}
\caption[]{
Constraints on the size of X-ray emitting shocked region and pre-shock number 
density for V1647 Ori from measured X-ray luminosities. The dotted and dashed 
line refer to the HIGH and LOW state of X-ray emission, respectively.
}
\end{center}
\end {figure*}

\end{document}